\documentclass[english,aip,jap,preprint,superscriptaddress,showpacs,endfloats*]{revtex4-1}
\usepackage[T1]{fontenc}
\usepackage[latin9]{inputenc}
\usepackage{color}
\usepackage{babel}
\usepackage{units}
\usepackage{ifsym}
\usepackage{amstext}
\usepackage{amssymb}
\usepackage{wasysym}
\usepackage{graphicx}
\usepackage[unicode=true,
 bookmarks=true,bookmarksnumbered=false,bookmarksopen=false,
 breaklinks=false,pdfborder={0 0 0},backref=false,colorlinks=true]
 {hyperref}
\hypersetup{
 pdfauthor={V. Zaretskey},
 urlcolor=blue,citecolor=red,bookmarksopen=true,bookmarksopenlevel=1,pdfstartview=FitH}
\usepackage{breakurl}

\makeatletter

\DeclareRobustCommand{\greektext}{%
  \fontencoding{LGR}\selectfont\def\encodingdefault{LGR}}
\DeclareRobustCommand{\textgreek}[1]{\leavevmode{\greektext #1}}
\DeclareFontEncoding{LGR}{}{}
\DeclareTextSymbol{\~}{LGR}{126}
\newcommand{\lyxmathsym}[1]{\ifmmode\begingroup\def\b@ld{bold}
  \text{\ifx\math@version\b@ld\bfseries\fi#1}\endgroup\else#1\fi}

\providecommand{\tabularnewline}{\\}

 \@ifundefined{textcolor}{}
 {%
   \definecolor{BLACK}{gray}{0}
   \definecolor{WHITE}{gray}{1}
   \definecolor{RED}{rgb}{1,0,0}
   \definecolor{GREEN}{rgb}{0,1,0}
   \definecolor{BLUE}{rgb}{0,0,1}
   \definecolor{CYAN}{cmyk}{1,0,0,0}
   \definecolor{MAGENTA}{cmyk}{0,1,0,0}
   \definecolor{YELLOW}{cmyk}{0,0,1,0}
 }

\usepackage{epstopdf}

\makeatother

\begin{document}

\title{Decoherence in a Pair of Long-Lived Cooper-Pair Boxes}

\date{\today}

\author{V. Zaretskey}

\author{S. Novikov}

\author{B. Suri}

\affiliation{Department of Physics, University of Maryland, College Park, Maryland,
20742}

\affiliation{Laboratory for Physical Sciences, College Park, Maryland, 20740}

\author{Z. Kim}

\affiliation{Department of Physics, University of Maryland, College Park, Maryland,
20742}

\author{F. C. Wellstood}

\affiliation{Department of Physics, University of Maryland, College Park, Maryland,
20742}

\affiliation{Joint Quantum Institute and Center for Nanophysics and Advanced Materials,
College Park, Maryland, 20742}

\author{B. S. Palmer}

\affiliation{Laboratory for Physical Sciences, College Park, Maryland, 20740}
\begin{abstract}
We have investigated the decoherence of quantum states in two $\mbox{Al}/\mbox{AlO}_{x}/\mbox{Al}$
Cooper-pair boxes coupled to lumped element superconducting LC resonators.
At $\unit[25]{mK}$, the first qubit had an energy relaxation time
$T_{1}$ that varied from $\unit[30]{\text{\textgreek{m}}s}$ to $\unit[200]{\text{\textgreek{m}}s}$
between $4$ and $\unit[8]{GHz}$ and displayed an inverse correlation
between $T_{1}$ and the coupling to the microwave drive line. The
Ramsey fringe decay times $T_{2}^{*}$ were in the $\unit[200-500]{ns}$
range while the spin echo envelope decay times $T_{echo}$ varied
from $\unit[2.4-3.3]{\lyxmathsym{\textgreek{m}}s}$, consistent with
$1/f$ charge noise with a high frequency cutoff of $\unit[0.2]{MHz}$.
A second Cooper-pair box qubit with similar parameters showed $T_{1}=\unit[4-30]{\text{\textgreek{m}}s}$
between $\unit[4-7.3]{GHz}$, and that the $T_{1}$ and the coupling
were again inversely correlated. Although the lifetime of the second
device was shorter than that of the first device, the dependence on
coupling in both devices suggests that further reduction in coupling
should lead to improved qubit performance.
\end{abstract}


\maketitle

\section*{Introduction}

There has been great progress in increasing the relaxation times of
superconducting qubits over the last decade. Many devices have been
reported with relaxation times in the $\unit[1-10]{\lyxmathsym{\textgreek{m}}s}$
range and more recently devices have started to appear with relaxation
times of $\unit[100]{\lyxmathsym{\textgreek{m}}s}$ or greater.\cite{houck2008controlling,bertet2005dephasing,manucharyan2012evidence,paik2011observation}
Qubit relaxation has been attributed to a variety of different physical
mechanisms including dielectric loss,\cite{martinis2005decoherence}
coupling to discrete two level fluctuators,\cite{kim2008anomalous,simmonds2004decoherence}
excess non-equilibrium quasiparticles,\cite{martinis2009energydecay,lenander2011measurement,catelani2011quasiparticle}
and coupling to lossy electromagnetic modes.\cite{houck2008controlling}
Recent work that addresses some of these underlying causes of energy
loss include the use of absorbers and shielding to block infrared
radiation responsible for non-equilibrium quasiparticles,\cite{barends2011minimizing,corcoles2011protecting}
decoupling the qubit from the environment,\cite{kim2011decoupling}
and placing the qubit in a 3D resonator to both decouple it and reduce
the contribution of lossy materials.\cite{paik2011observation}

We recently reported on a Cooper-pair box (CPB) qubit with an excited
state lifetime $T_{1}\approx\unit[200]{\mu s}$, about one order of
magnitude larger than typically reported.\cite{kim2011decoupling}
This device showed a correlation between the relaxation time $T_{1}$
and the magnitude of the decoupling between the device and the microwave
drive line. Here we address some key questions that naturally follow
from these results. First, although the excited state lifetime provides
a measure of the high frequency noise affecting the qubit,\cite{schoelkopf2002qubitsas}
what is the character of the low frequency noise that is responsible
for dephasing?\cite{ithier2005decoherence,martinis2003decoherence}
Moreover, is there any correspondence between the low frequency and
high frequency noise? Second, given the significant improvement in
lifetime, how reproducible are these results? We address the first
issue by measuring Ramsey fringes, Rabi oscillations, and a spin echo
experiment. To understand the reproducibility of these results, we
fabricated a second qubit based on the original design and measured
its lifetime and coupling.

\section*{Theoretical Description}

\begin{figure}
\includegraphics[width=1\columnwidth]{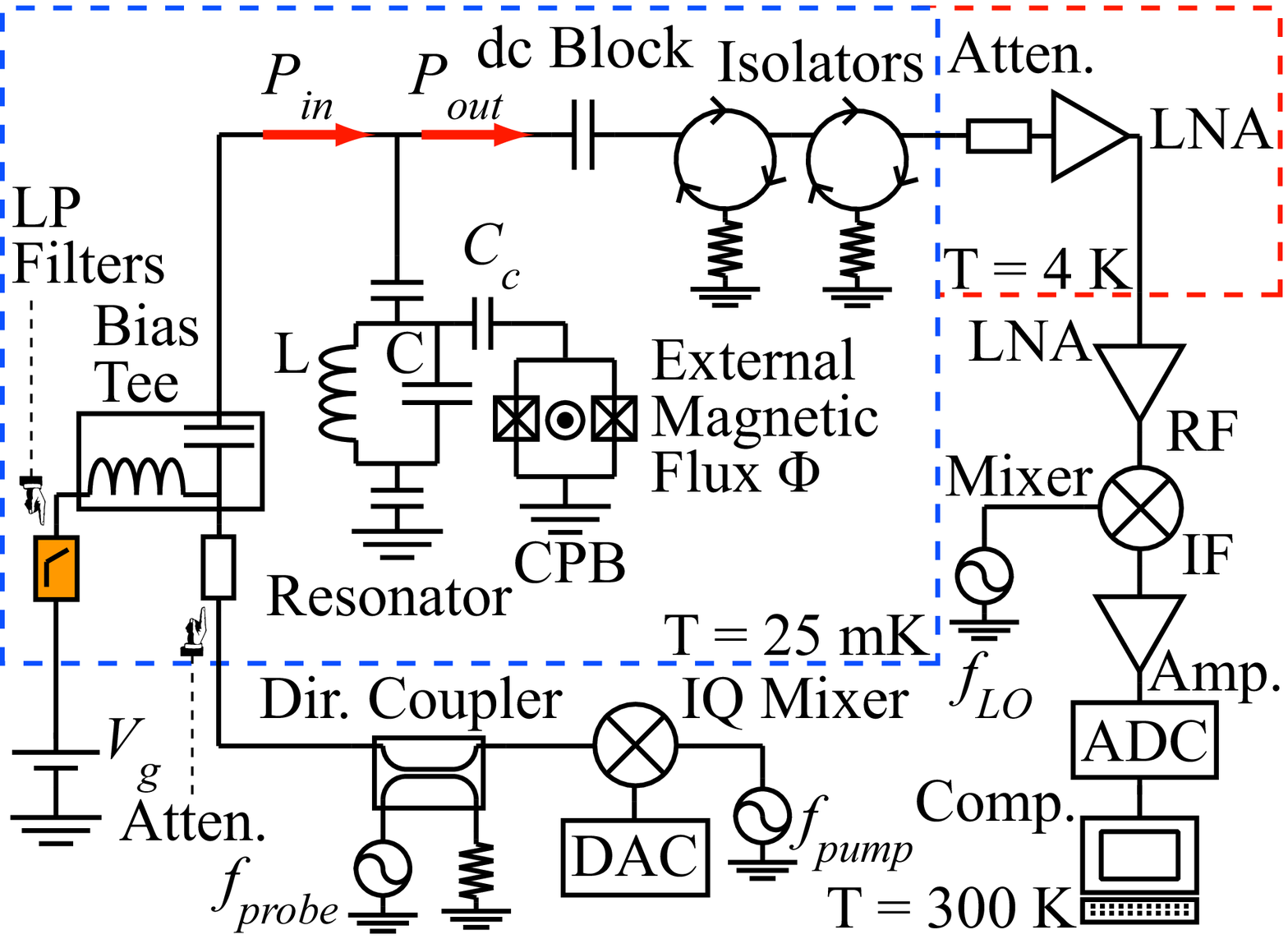}

\caption{Simplified schematic of CPB, resonator, and experimental setup. The
CPB is coupled through capacitor $C_{c}$ to a lumped element LC resonator.
The qubit state is read out via a coherent heterodyne measurement
of the transmitted microwave signal at frequency $f_{probe}$ which
is amplified, mixed with a local oscillator at frequency $f_{LO}$
and finally digitized. The CPB transition frequency is controlled
by the gate voltage $V_{g}$ and an external magnetic flux $\Phi$
and the CPB state is coherently manipulated using shaped microwave
pulses at frequency $f_{pump}$.\label{fig:setup}}
\end{figure}
A Cooper-pair box\cite{bouchiat1998quantum} consists of a small superconducting
island connected to a superconducting reservoir (ground) through two
ultrasmall Josephson tunnel junctions with total critical current
$I_{0}$ and total junction capacitance $C_{J}$ {[}see Fig. \ref{fig:setup}{]}.
A gate electrode with capacitance $C_{g}$ to the island allows us
to apply a bias voltage $V_{g}$ and control the system's electrostatic
energy, while an external flux $\Phi$ though the superconducting
loop can be used to adjust the critical current $I_{0}$. The island
has a total capacitance to ground of $C_{\Sigma}=C_{J}+C_{g}$ which
sets the charging energy $E_{c}=e^{2}/2C_{\Sigma}$. $E_{c}$ and
the Josephson energy $E_{J}=\hbar I_{0}/2e$ form two competing energy
scales that determine the optimal quantization basis, the sensitivity
to various types of noise and the operating parameters. In the limit
$E_{c}\gg E_{J}$, the CPB Hamiltonian yields highly anharmonic energy
levels. For our purpose, only the two lowest levels need to be considered
and we can write 
\begin{equation}
H_{CPB}\cong\frac{\hbar\omega_{q}}{2}\sigma_{z}\label{eq:hcpb}
\end{equation}
where $\sigma_{z}$ is the Pauli spin operator,
\begin{equation}
\hbar\omega_{q}=\sqrt{\left[4E_{c}\left(1-n_{g}\right)\right]^{2}+E_{J}^{2}}\label{eq:cpb-parabola}
\end{equation}
is the ground to first excited state transition energy, and $n_{g}=C_{g}V_{g}/e$
is the reduced gate voltage.

To read out the state of the qubit, we coupled our qubit to a quasi-lumped
element inductor-capacitor (LC) resonator that was in turn coupled
to a transmission line {[}see Fig. \ref{fig:images}(a){]}. We probed
the LC resonance frequency by applying microwave power and recording
the transmitted microwave signal. This is a dispersive readout in
which the qubit produces a state-dependent reactance that perturbs
the resonance frequency of the resonator. For weak qubit-resonator
coupling $g$ and large detuning $\Delta=\omega_{q}-\omega_{r}$ between
the qubit transition frequency $\omega_{q}$ and the resonator resonance
frequency $\omega_{r}$, the system Hamiltonian is approximately\cite{wallraff2005approaching,blais2004cavityquantum}
\begin{equation}
H=\hbar\left(\omega_{r}+\frac{g^{2}}{\Delta}\sigma_{z}\right)\left(a^{\dagger}a+\frac{1}{2}\right)+\frac{\hbar\omega_{q}}{2}\sigma_{z}\label{eq:htotal}
\end{equation}
where
\begin{equation}
\hbar g=\left(2E_{c}C_{c}/e\right)\sqrt{\hbar\omega_{r}/2C}
\end{equation}
is the qubit-resonator interaction strength, $C_{c}$ is the coupling
capacitance between the resonator and the island of the CPB and $C$
is the capacitance of the resonator {[}see Fig. \ref{fig:setup}{]}.
Equation \ref{eq:htotal} implies that the bare resonator frequency
$\omega_{r}$ is dispersively shifted by $\chi=\pm g^{2}/\Delta$
depending on the state of the qubit. If $\chi\ll\kappa$, where $\kappa=\omega_{r}/Q_{L}$
is the resonator linewidth and $Q_{L}$ is the resonator quality factor,
the average phase of the transmitted signal at $\omega_{r}$ is linearly
dependent on the excited state occupation probability. On the other
hand if $\chi\gg\kappa$, then when on-resonance microwave power is
applied, the average in-phase or quadrature transmitted voltage is
proportional to the excited state occupation probability.\cite{gambetta2007protocols}

Qubit decoherence is caused by relaxation and dephasing. The relaxation
rate can be found using Fermi's golden rule and depends on the spectral
density of noise at the transition frequency and the transition matrix
element.\cite{schoelkopf2002qubitsas,astafiev2004quantum} Sources
of noise may be external, such as thermal or instrumentation noise
propagating down imperfectly filtered control lines, or local to the
qubit, such as nearby lossy materials. If multiple uncorrelated noise
channels are present then the total relaxation rate is given by the
sum of the individual decay rates.

For the Cooper-pair box, a key factor that governs the relaxation
time is the sensitivity to voltage or charge perturbations. At the
CPB sweet spot ($n_{g}=1$), small changes $\Delta V_{g}$ in the
gate voltage produce a relaxation rate via a matrix element:
\begin{equation}
\left\langle \vphantom{\frac{\partial\hat{H}}{\partial V_{g}}}g\right|\frac{\partial\hat{H}}{\partial V_{g}}\left|\vphantom{\frac{\partial\hat{H}}{\partial V_{g}}}e\right\rangle =\frac{eC_{g}E_{J}}{\hbar\omega_{q}C_{\Sigma}}=\frac{2C_{g}E_{c}}{e}\label{eq:chrgtrns}
\end{equation}
where $\left|g\right\rangle $ and $\left|e\right\rangle $ are the
ground state and excited state of the CPB, respectively.

In contrast, dephasing is caused by adiabatic changes in the transition
frequency between $\left|g\right\rangle $ and $\left|e\right\rangle $
and depends on low frequency noise components, the sensitivity of
the transition frequency to external parameters, and the how the qubit
state is manipulated.\cite{ithier2005decoherence,martinis2003decoherence}
The first order sensitivity of the CPB $\left|g\right\rangle $ to
$\left|e\right\rangle $ transition frequency to low frequency charge
noise is
\begin{equation}
\frac{\partial\omega_{q}}{\partial n_{g}}=\frac{1}{\hbar^{2}}\frac{\left(4E_{c}\right)^{2}\left(n_{g}-1\right)}{\omega_{q}}\mbox{,}\label{eq:1stchrgsens}
\end{equation}
which vanishes at the sweet spot. The second order sensitivity at
the sweet spot is
\begin{equation}
\frac{\partial^{2}\omega_{q}}{\partial n_{g}^{2}}=\frac{1}{\hbar^{2}}\frac{\left(4E_{c}\right)^{2}}{\omega_{q}}{\textstyle .}\label{eq:2ndchrgsens}
\end{equation}
Voltage and charge fluctuations are the dominant types of noise affecting
our qubit and we focus our discussion there. Other noise types of
noise, such as flux noise,\cite{wellstood1987lowfrequency} should
also be present but their effect should be much smaller than charge
noise for our design. Assuming typical values of flux noise, charge
noise should dominate even at second order.\cite{ithier2005decoherence,schuster2007circuit}

\section*{Experimental Details}

\begin{figure}
\includegraphics[width=1\columnwidth]{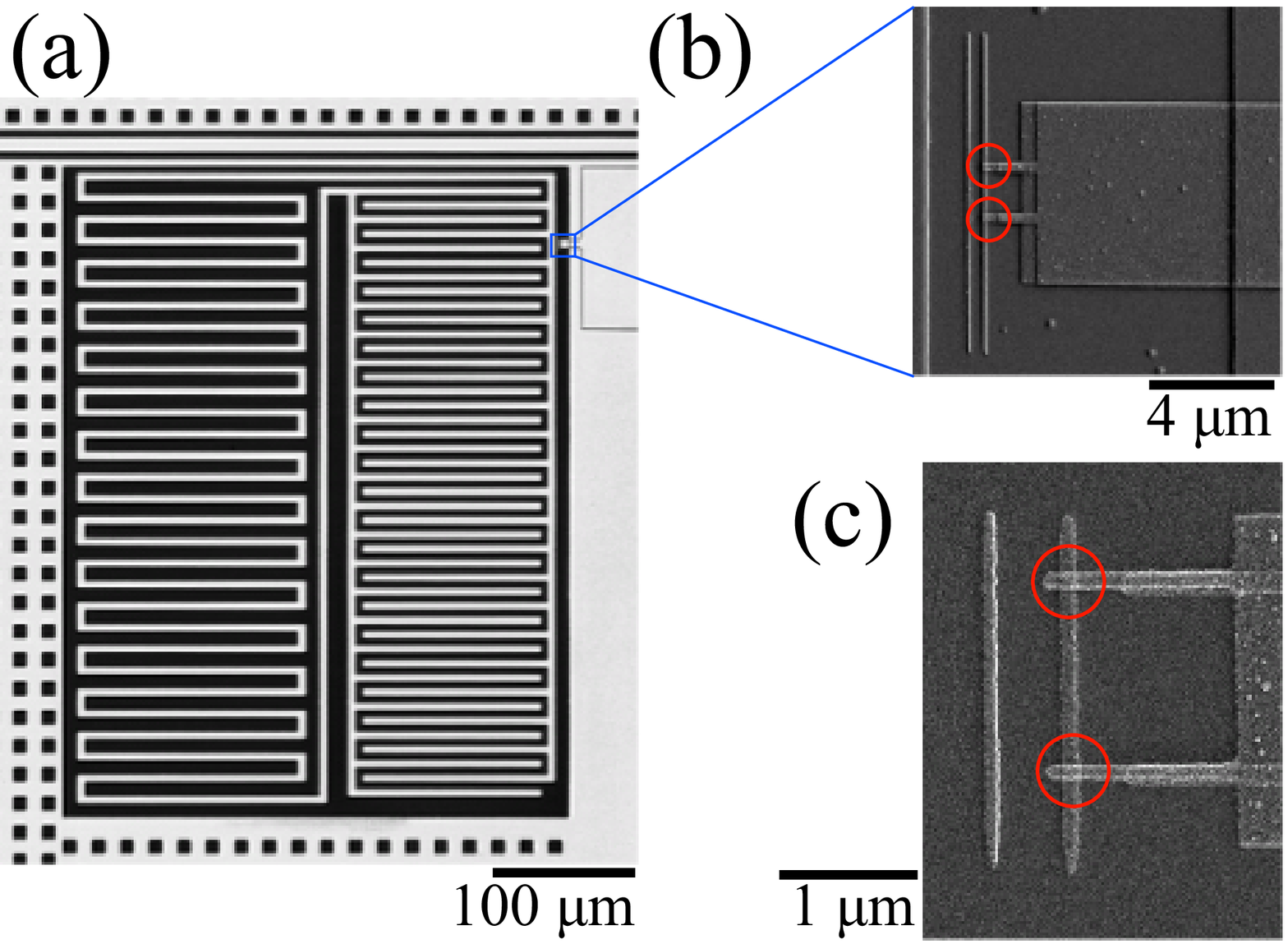}

\caption{(a) Optical image showing the lumped element resonator coupled to
a coplanar waveguide (CPW) transmission line and surrounded by a perforated
ground plane. Light regions are aluminum metalization and dark are
sapphire substrate. (b) Scanning electron microscope image of device
2 CPB, located between the resonator interdigital capacitor and ground
plane. The twinned features are a consequence of double-angle evaporation.
Josephson junctions are marked with red circles. (c) Scanning electron
microscope image of device 1 CPB.\label{fig:images}}
\end{figure}
We fabricated high quality factor superconducting resonators using
standard photolithography and lift-off techniques. Each resonator
was made from $\unit[100]{nm}$ thick films of thermally evaporated
Al on a c-plane sapphire wafer. The lumped element resonators consisted
of a meander inductor $L\approx\unit[2]{nH}$ and an interdigital
capacitor $C\approx\unit[400]{fF}$ coupled to a coplanar waveguide
transmission line {[}see Fig. \ref{fig:images}(a){]}. The resonance
frequency of the first device\cite{kim2011decoupling} was $\omega_{r,1}/2\pi=\unit[5.44]{GHz}$
with a loaded quality factor $Q_{L,1}=22,000$, external quality factor
$Q_{e,1}=70,000$ and internal quality factor $Q_{i,1}=32,000$ at
low power. The second device had a very similar layout and a nearly
identical resonance frequency of $\omega_{r,2}/2\pi=\unit[5.47]{GHz}$,
but the loaded quality factor was $Q_{L,2}=35,000$, the external
quality factor was $Q_{e,2}=47,000$ and the internal quality factor
was $Q_{i,2}=147,000$. Device 2 was designed with a lower external
quality factor than that of device 1 to increase the bandwidth available
during readout. The discrepancy in the internal quality factors between
the two devices was likely due to variation in the fabrication, for
example in the quality and thickness of the native Al oxide, the roughness
of the surface of the Al, or the quality of the interface between
the evaporated Al and the sapphire wafer.

\begin{table}
\caption{Summary of the qubit and resonator parameters, decoherence, and noise
bounds for devices 1 and 2.\label{tab:device-summary}}
\begin{tabular}{lllll}
\hline 
 &  & Device 1 &  & Device 2\tabularnewline
\cline{3-5} 
 &  &  &  & \tabularnewline
\hline 
$\omega_{r}/2\pi$ & \enskip{} & $\unit[5.446]{GHz}$ & \enskip{} & $\unit[5.472]{GHz}$\tabularnewline
$Q_{L}$ &  & $22,000$ &  & $35,000$\tabularnewline
$Q_{i}$ &  & $32,000$ &  & $147,000$\tabularnewline
$Q_{e}$ &  & $70,000$ &  & $47,000$\tabularnewline
 &  &  &  & \tabularnewline
\hline 
$\omega_{a}/2\pi$ &  & $\unit[4-8]{GHz}$ &  & $\unit[4-7.3]{GHz}$\tabularnewline
$E_{J,max}/h$ & \enskip{} & $\unit[19]{GHz}$ & \enskip{} & $\unit[7.33]{GHz}$\tabularnewline
$E_{c}/h$ &  & $\unit[6.24]{GHz}$ &  & $\unit[4.3]{GHz}$\tabularnewline
$C_{g}$ &  & $\unit[4.5]{aF}$ &  & $\unit[19.1]{aF}$\tabularnewline
$g/2\pi$ &  & $\unit[5]{MHz}$ &  & $\unit[10-15]{MHz}$\tabularnewline
 &  &  &  & \tabularnewline
\hline 
$T_{1}$ & \enskip{} & $\unit[30-200]{\lyxmathsym{\textgreek{m}}s}$ & \enskip{} & $\unit[4-30]{\lyxmathsym{\textgreek{m}}s}$\tabularnewline
$T_{2}^{*}$ &  & $\unit[200-500]{ns}$ &  & $\unit[60]{ns}$\tabularnewline
$T_{echo}$ &  & $\unit[2.4-3.3]{\lyxmathsym{\textgreek{m}}s}$ &  & ---\tabularnewline
$T^{\prime}$ &  & $\unit[1-2]{\lyxmathsym{\textgreek{m}}s}$ &  & $\unit[0.2-1.8]{\lyxmathsym{\textgreek{m}}s}$\tabularnewline
$S_{q}\left(f=\unit[1]{Hz}\right)$ &  & $\unit[\left(3\times10^{-3}\right)^{2}]{\textit{e}^{2}/Hz}$ &  & $\unit[\left(1\times10^{-2}\right)^{2}]{\textit{e}^{2}/Hz}$\tabularnewline
$S_{q}\left(f=\unit[4.5]{GHz}\right)$ &  & $\unit[10^{-18}]{\textit{e}^{2}/Hz}$ &  & $\unit[10^{-17}]{\textit{e}^{2}/Hz}$\tabularnewline
\hline 
\end{tabular}
\end{table}
The CPB was subsequently defined by e-beam lithography. We used a
bilayer stack of MMA(8.5)MAA copolymer and ZEP520A e-beam resist to
facilitate lift-off and reduce proximity exposure during writing.
Al films were deposited using double-angle evaporation with an intermediate
thermal oxidation step to create the Josephson tunnel junctions;\cite{dolan1977offsetmasks}
$\unit[30]{nm}$ thick Al island and $\unit[50]{nm}$ thick Al leads
were deposited in an e-beam evaporator {[}see Fig. \ref{fig:images}(b,
c){]}. We set the charging energy $E_{c}$ and Josephson energy $E_{J}$
by choosing the lithographically defined junction size and the oxygen
exposure. Device 1 had $E_{c,1}/h=\unit[6.24]{GHz}$ and $E_{J,max,1}/h=\unit[19]{GHz}$
and $E_{J}$ was tuned with an external magnetic field to the $\unit[4-8]{GHz}$
range. Device 2 was designed with a smaller $E_{c}$ to reduce sensitivity
to charge noise {[}see Eq. (\ref{eq:2ndchrgsens}){]} and a smaller
$E_{J,max}$ to enable operation at the double sweet spot ($n_{g}=1$
and no external flux bias) if desired. It had $E_{c,2}/h=\unit[4.3]{GHz}$,
$E_{J,max,2}/h=\unit[7.33]{GHz}$, and $E_{J}/h$ could be tuned as
low as $\unit[4]{GHz}$. Table \ref{tab:device-summary} summarizes
the parameters of the two devices and their resonators.

Each sample was mounted in a rf-tight Cu box with Al wire bond connections
to the chip. The sample box was anchored to the mixing chamber of
an Oxford Instruments 100 dilution refrigerator with a base temperature
of $\unit[25]{mK}$. We used cold attenuators on the input line at
$\unit[4.2]{K}$ ($\unit[10]{dB}$), $\unit[0.6]{K}$ ($\unit[20]{dB}$)
and $\unit[25]{mK}$ ($\unit[30]{dB}$) and two $\unit[18]{dB}$ isolators
on the output line at $\unit[25]{mK}$ to filter thermal noise from
higher temperatures. A filtered dc bias line for applying the gate
voltage $V_{g}$ was coupled to the input line using a bias tee before
the device and a dc block was placed after the sample box {[}see Fig.
\ref{fig:setup}{]}. The output microwave signal was amplified with
a HEMT amplifier\cite{weinreb} sitting in the He bath.

Spectroscopic and lifetime measurements were performed by continuously
monitoring the transmitted amplitude and phase at the resonator's
resonance frequency using a weak microwave drive. For qubit spectroscopy,
the gate voltage was swept adiabatically while a second continuous
microwave tone was stepped in frequency to excite the qubit.

\begin{figure*}
\includegraphics[width=1\textwidth]{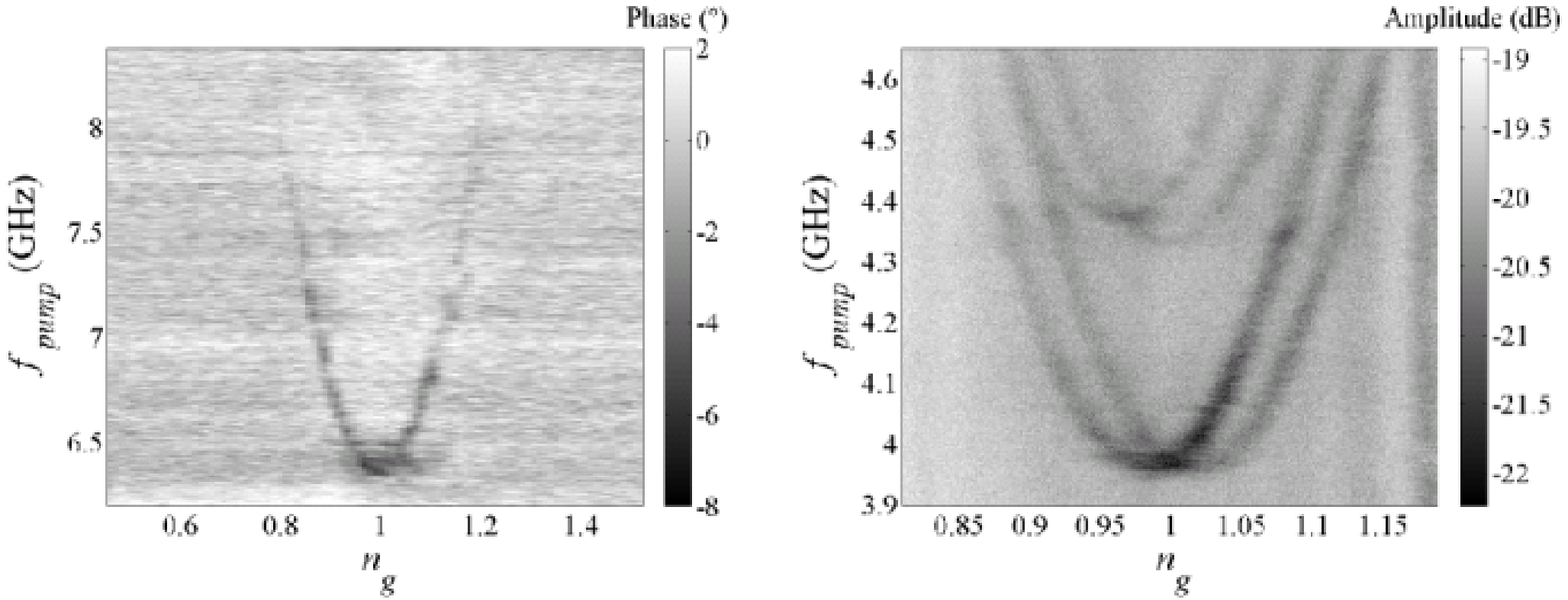}

\caption{(a) Measured transition spectrum of CPB device 1.\cite{kim2011decoupling}
The grayscale density plot shows the change in the phase of the transmitted
microwave probe signal as a function of the reduced gate voltage $n_{g}$
and pump tone $f_{pump}$. Darker indicates more absorption. (b) Measured
spectrum of device 2 displayed as the amplitude of the transmitted
probe signal.\cite{zaretskey2013spectroscopy} Four parabolas are
clearly visible.\label{fig:spectrum}}
\end{figure*}
Initial characterization of device 1 has been reported by Kim, \textit{et
al.}\cite{kim2011decoupling} Figure \ref{fig:spectrum}(a) shows
a grayscale plot of the typical transmitted signal phase as functions
of $n_{g}$ and pump frequency $f_{pump}$. The data closely resembled
a parabola and allowed us to extract $E_{c}$ and $E_{J}$ by fitting
to Eq. \ref{eq:cpb-parabola}.

To avoid dephasing effects due to the back-action of the probe photons\cite{schuster2005acstark,blais2004cavityquantum,wallraff2005approaching}
for Rabi oscillation, Ramsey fringes, and spin echo measurements,
we probed the resonator at $\omega_{r}$ only after the completion
of any qubit pulse sequence. Qubit measurements were repeated every
$\unit[0.2-2]{ms}$, a delay equal to at least several lifetimes $T_{1}$
to allow the qubit to relax to the ground state, and averaged $5000-10,000$
times. We used a coherent heterodyne setup to record the phase and
amplitude of the transmitted $\omega_{r}$ signal at $\unit[500]{ns}$
time bins. Specifically, the signal from the HEMT was amplified at
room temperature, mixed with a local oscillator tone to an intermediate
frequency of $\unit[2]{MHz}$ and then digitally sampled at a typical
sampling rate of $\unit[20]{MSa/s}$. A second reference tone was
split off from the probe signal source and similarly mixed and digitally
sampled. Both signals were then passed through a second stage of software
demodulation. Manipulation of the qubit state was performed with a
separate pulse shaping system consisting of a two-channel $\unit[1]{GSa/s}$
DAC board\cite{martinis} that supplied control voltages to a $\unit[4-8]{GHz}$
IQ mixer {[}see Fig. \ref{fig:setup}{]}. This allowed us to produce
microwave pulses at frequency $\omega_{q}$ with a $\unit[3]{ns}$
Gaussian rise time and arbitrary envelope and phase to perform the
necessary qubit manipulations. All components were locked to a $\unit[10]{MHz}$
Rb atomic clock.\cite{srs}

\section*{Lifetime and Dephasing Studies of Device 1}

\begin{figure}
\includegraphics[width=1\columnwidth]{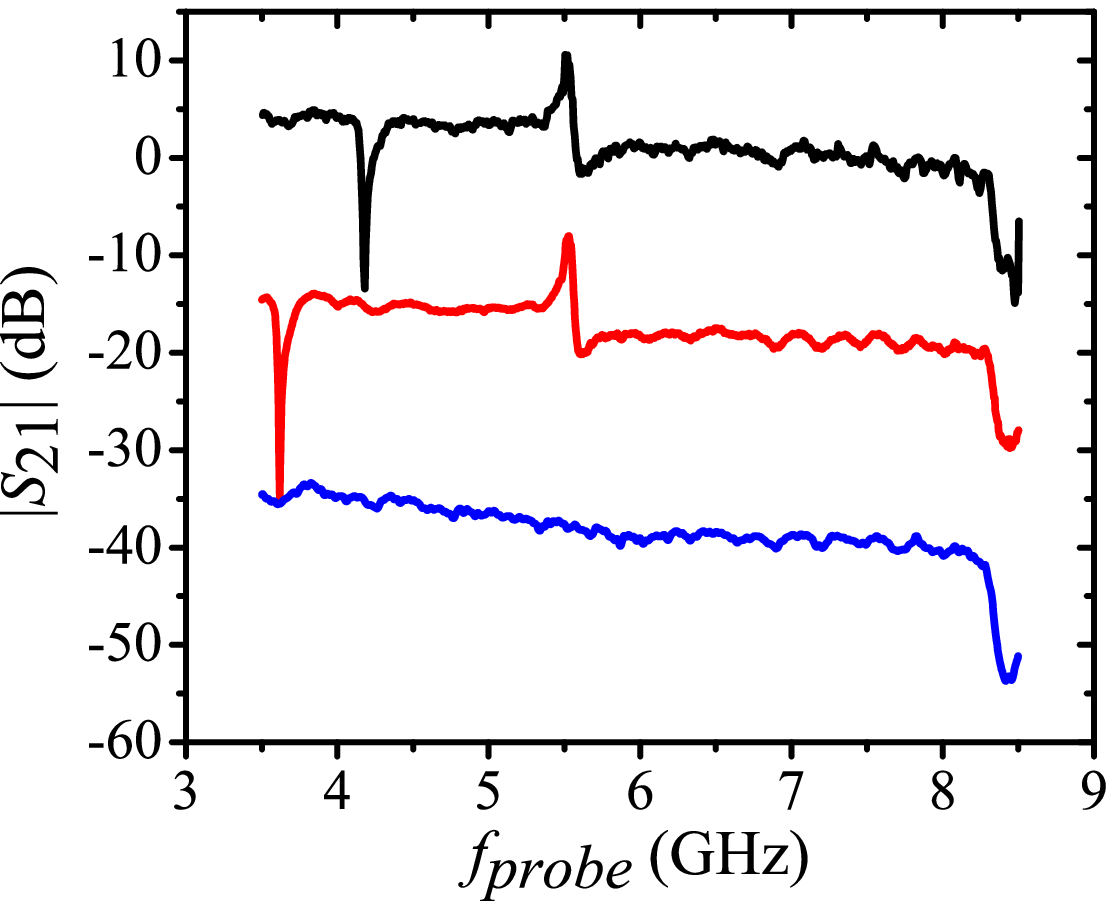}

\caption{History of the transmitted voltage ratio $\left|S_{21}\right|$ through
the experimental setup {[}see Fig. \ref{fig:setup}{]}. Top curve
(black) is transmission amplitude $\left|S_{21}\right|$ during the
lifetime and decoupling characterization of device 1 carried out by
Z. Kim, \textit{et al.}\cite{kim2011decoupling} We recorded the middle
curve (red) after modifying the grounds of the CPW launchers. This
shifted the box resonance from $\unit[4.2]{GHz}$ to $\unit[3.5]{GHz}$
and was the situation during our dephasing measurements of device
1. Afterward, we installed a different cryogenic amplifier and placed
device 2 in a different sample box. As seen in the bottom curve (blue),
this removed both the box resonance and the self-resonance of the
amplifier at $\unit[5.67]{GHz}$. Successive curves have been offset
by $\unit[20]{dB}$ for clarity.\label{fig:s21-history}}
\end{figure}
After the spectroscopy shown in Fig. \ref{fig:spectrum}(a) was taken,
the experimental setup was modified by changing the grounds in the
coplanar waveguide launchers that connect the chip to the microwave
lines. This change moved a spurious resonance due to device packaging
from $\unit[4.2]{GHz}$ down to $\unit[3.5]{GHz}$ {[}see Fig. \ref{fig:s21-history}{]}.
The lifetime and Rabi coupling at $n_{g}=1$ was remeasured after
tuning $E_{J}/h=\unit[4.5]{GHz}$, the frequency where the lifetime
was longest and coupling smallest in the previous cooldown.\cite{kim2011decoupling}
The coupling between the qubit and the microwave line increased from
$\unit[0.13]{MHz/\mu V}$ before the change to $\unit[0.53]{MHz/\mu V}$
after and there was a corresponding decrease in the lifetime $T_{1}$
from $\unit[205]{\mu s}$ to $\unit[61]{\mu s}$. Assuming that this
relaxation is due to charge noise, the power spectral density of charge
noise $S_{Q}\left(f\right)$ and the lifetime $T_{1}$ are related
by\cite{astafiev2004quantum,schoelkopf2002qubitsas}

\begin{equation}
S_{Q}\left(f_{q}\right)=\left(\frac{e\hbar}{2E_{c}}\right)^{2}\frac{1}{T_{1}}\label{eq:sqandt1}
\end{equation}
and $T_{1}=\unit[61]{\mu s}$ leads to a bound on the charge noise
of $S_{Q}\left(f=\unit[4.5]{GHz}\right)\leq3\times\unit[10^{-18}]{\textit{e}^{2}/Hz}$.\cite{astafiev2004quantum,schoelkopf2002qubitsas}

\begin{figure*}
\includegraphics[width=1\textwidth]{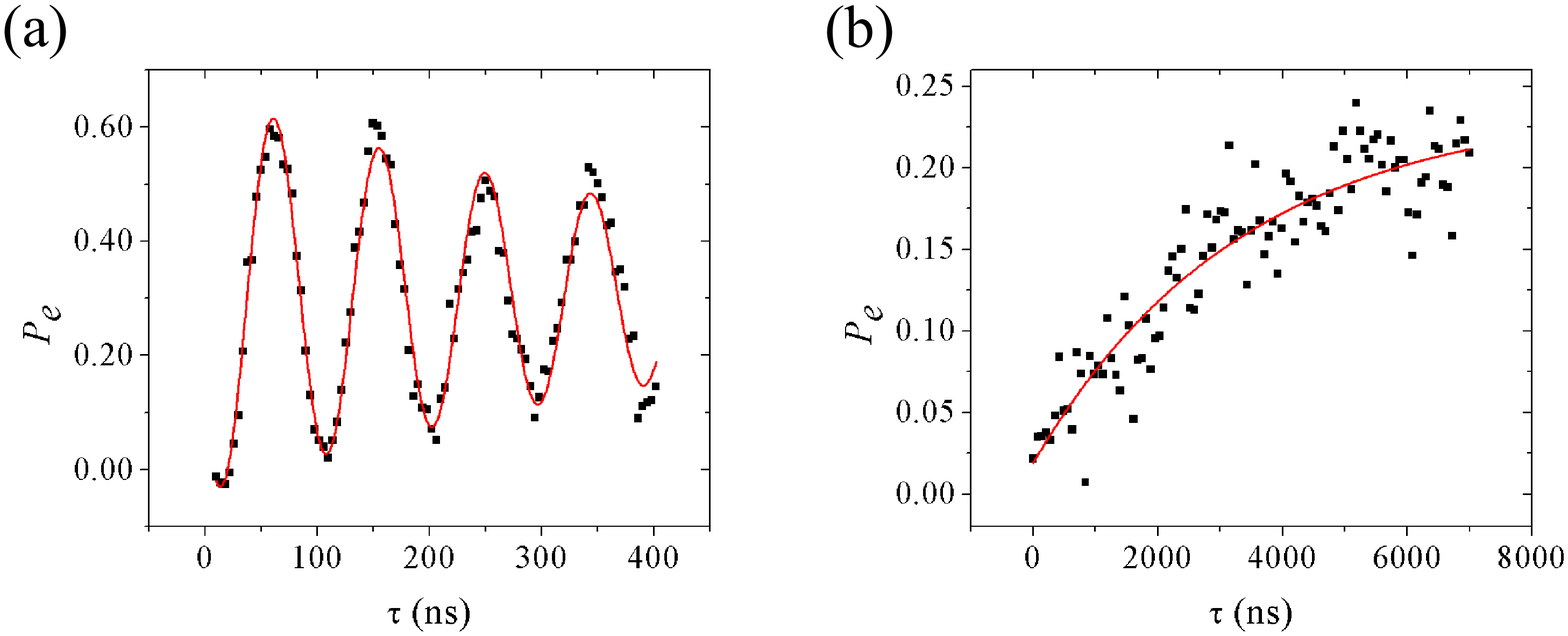}

\caption{(a) Device 1 Ramsey fringes observed in the qubit population $P_{e}$
versus the time interval $\tau$ between two $\pi/2$ pulses. Filled
squares are measured data and the red curve is a fit to an exponentially
damped sinusoid with decay time constant $T_{2}^{*}=\unit[500]{ns}$
and a detuning of $\unit[10.6]{MHz}$. (b) Qubit population $P_{e}$
after a spin echo pulse sequence consisting of $\pi/2$ and $-\pi/2$
pulses separated by a delay $\tau$ and an intervening $\pi$ pulse
at time $\tau/2$. Filled squares show the data while the red curve
is a fit to an exponential decay with time constant $T_{echo}=\unit[3.3]{\text{\textgreek{m}}s}$.\label{fig:dephasing}}
\end{figure*}
We also measured the dephasing of device 1 during free evolution by
performing Ramsey fringe experiments and during driven evolution by
performing Rabi oscillations. From driving Rabi oscillations at $\omega_{q}/2\pi=\unit[5.949]{GHz}$
we obtained a Rabi decay time $T^{\prime}=\unit[1-2]{\mu s}$ and
calibrated the drive duration to produce a $\pi/2$ pulse in $\unit[26.0]{ns}$.
For Ramsey oscillations, we then applied two phase coherent $\pi/2$
pulses separated by a delay $\tau$ and at frequency $\omega_{pump}=\omega_{q}+\Delta\omega$,
i.e. detuned by $\Delta\omega$ from the qubit transition. The excited
state probability after the second $\pi/2$ pulse displayed decaying
oscillations of frequency $\Delta\omega/2\pi$ and the envelope of
the oscillations yields the Ramsey decay time $T_{2}^{*}$. The decay
time is sensitive to both homogeneous and inhomogeneous broadening.
Our observed $T_{2}^{*}$ were in the $\unit[200-500]{ns}$ range
{[}see Fig. \ref{fig:dephasing}(a){]}. Assuming charge noise with
a $1/f$ spectrum is the dominant noise source, from Eq. (\ref{eq:2ndchrgsens})
and the dependence of Ramsey fringes decay time on $1/f$ noise,\cite{ithier2005decoherence,martinis2003decoherence}
we extract a bound on the noise amplitude of $S_{Q}\left(f=\unit[1]{Hz}\right)\leq\unit[\left(3\times10^{-3}\right)^{2}]{\textit{e}^{2}/Hz}$.
This value is similar to that reported for other charge qubits\cite{wallraff2005approaching,siddiqi2006dispersive,vion2002manipulating,metcalfe2007measuring}
and is about an order of magnitude larger than low frequency charge
noise observed in $\mbox{Al}/\mbox{AlO}_{x}/\mbox{Al}$ single-electron
transistors.\cite{kenyon2000temperature,zimmerli1992noisein} If we
extrapolate this noise to $\unit[4.5]{GHz}$ the result is $S_{Q}\left(f=\unit[4.5]{GHz}\right)\leq\unit[2\times10^{-15}]{\textit{e}^{2}/Hz}$.
This is three orders of magnitude larger than that extracted from
the $T_{1}$ data\cite{kim2011decoupling} on device 1, suggesting
that the charge noise spectrum cannot be scaling as $1/f$ into the
GHz range.

We further characterized the noise affecting the qubit with a spin
echo experiment. For this measurement, the following pulse sequence
was used: apply $\pi/2$ pulse, wait time $\tau/2$, apply an out
of phase $\pi$ pulse, wait time $\tau/2$, apply a second in-phase
$-\pi/2$ pulse, and finally measure the excited state probability.
The intervening $\pi$ pulse serves to refocus the phase and greatly
reduces the impact of low frequency noise. The excited state probability
decayed exponentially with time constant $T_{echo}$ in the $\unit[2.4-3.3]{\lyxmathsym{\textgreek{m}}s}$
range {[}see Fig. \ref{fig:dephasing}(b){]}. We note that one expects
$T_{echo}/T_{2}^{*}\approx4.5$ for wideband $1/f$ noise.\cite{ithier2005decoherence,martinis2003decoherence}
Our observed ratio of $T_{echo}/T_{2}^{*}\approx6$ is greater than
this and consistent with a soft cutoff, such as a transition to $1/f^{2}$
falloff, of the $1/f$ noise at $f_{c}\approx\unit[0.2]{MHz}$. A
similar cutoff was reported by Ithier \textit{et al.}\cite{ithier2005decoherence}
at a frequency of $f_{c}\approx\unit[0.4]{MHz}$.

\section*{Lifetime and Decoupling of Device 2}

To better understand the reproducibility of our results, we fabricated
a second device with nominally the same layout and characterized it
using similar methodology. We placed device 2 in a different Cu sample
box and installed a new HEMT amplifier, in the process removing both
the box resonance and the self-resonance of the amplifier at $\unit[5.67]{GHz}$
{[}see Fig. \ref{fig:s21-history}{]}.

Table \ref{tab:device-summary} summarizes the parameters of device
2 while Fig. \ref{fig:spectrum}(b) shows a plot of the transition
spectrum. A striking difference between the two samples was immediately
evident. For device 2 we observed 4 parabolas with varying curvatures
and transition frequencies. This overall structure was stable over
the course of four months that the sample was cold and persisted as
we tuned the transition frequency from $\unit[4.0-7.3]{GHz}$. We
believe the observed spectrum is a signature of the CPB coherently
coupling to two defects that modulate the charge and critical current.\cite{zaretskey2013spectroscopy}
Particularly notable are the combined curvature and transition frequency
offsets. Previous work has found that qubit-defect interactions can
significantly degrade performance\cite{kim2008anomalous,sillanpaa2007coherent,simmonds2004decoherence,oh2006elimination,kline2009josephson}
or, in our experience, make the qubit inoperable.\cite{schuster2007circuit,kim2010dissipative}
However this was not the case for device 2, as we were able to measure
reasonable excited state lifetimes $T_{1}$ and record Rabi oscillations
for transitions to any of the parabolas. From fits to the spectra
we extracted $E_{c}/h=\unit[4.3]{GHz}$.\cite{zaretskey2013spectroscopy}

\begin{figure*}
\includegraphics[clip,width=1\textwidth]{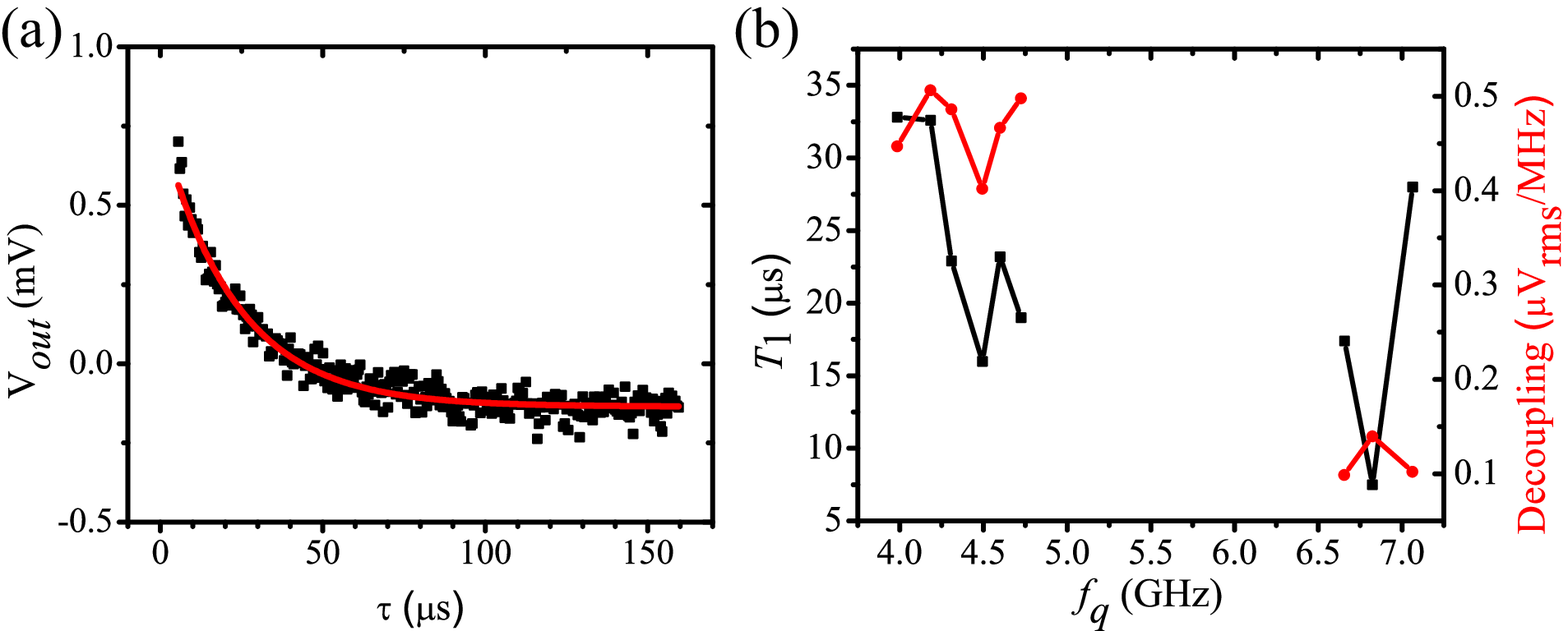}

\caption{(a) Energy relaxation measurement of device 2 showing in-phase transmitted
microwave amplitude versus time. The in-phase voltage $V_{out}$ is
proportional to the excited state probability $P_{e}$. The red curve
shows a fit to an exponential decay. (b) Correlation of the relaxation
time $T_{1}$ and qubit decoupling for device 2. Plotted are $T_{1}$
(\textifsymbol[ifgeo]{80}, left axis) and decoupling (\textcolor{red}{\CIRCLE{}},
right axis) versus CPB transition frequency $f_{q}$. The decoupling
is determined from the dependence of the Rabi frequency on the applied
pump power.\label{fig:t1}}
\end{figure*}
We measured the excited state lifetime by biasing the qubit at nominally
$n_{g}=1$ (at a minimum of one of the parabolas), applying a microwave
tone at $\omega_{q}$ to saturate the qubit, turning on the probe
tone, and then recording the transmitted signal. Figure \ref{fig:t1}(a)
shows a typical in-phase voltage trace and a fit to an exponential
decay. While we were able to operate the sample at any of the spectral
parabolas, one of the parabolas had the best measurement contrast.
When $E_{J}/\hbar$ was turned below $\omega_{r}$ this was the lowest
lying parabola (see parabola \#1 in Fig. \ref{fig:spectrum}(b)) while
for $E_{J}/\hbar>\omega_{r}$ it was the third highest (see parabola
\#3 in Fig. \ref{fig:spectrum}(b)).

In addition to measuring the lifetime, we also measured Rabi oscillations
and found $dV/df_{Rabi}$ where $f_{Rabi}$ is the frequency of Rabi
oscillations at the microwave drive rms amplitude $V$. This quantity
is a measure of the decoupling between the qubit and the microwave
drive. Rabi oscillations at various powers were recorded by pulsing
a microwave pump tone at $\omega_{q}$ for a fixed duration, turning
on the probe tone at $\omega_{r}$, and recording the transmitted
signal at $\omega_{r}$ during a $\unit[6-8]{\lyxmathsym{\textgreek{m}}s}$
window.\cite{blais2004cavityquantum,gambetta2007protocols} Typically
$5000-10,000$ such measurements were averaged at each pulse duration
and the process was repeated for a range of pump pulse durations.

Figure \ref{fig:t1}(b) shows a plot of the lifetime $T_{1}$ and
decoupling $dV/df_{Rabi}$ versus the CPB transition frequency for
both devices. For device 2 there is a gap in the data between $\unit[5.0-6.5]{GHz}$
where reduced visibility in all the parabolas prohibited data acquisition.\cite{zaretskey2013spectroscopy}
This loss of visibility was possibly due to a charged fluctuator that
was activated in this frequency range.

\section*{Comparison}

\begin{figure*}
\includegraphics[width=1\textwidth]{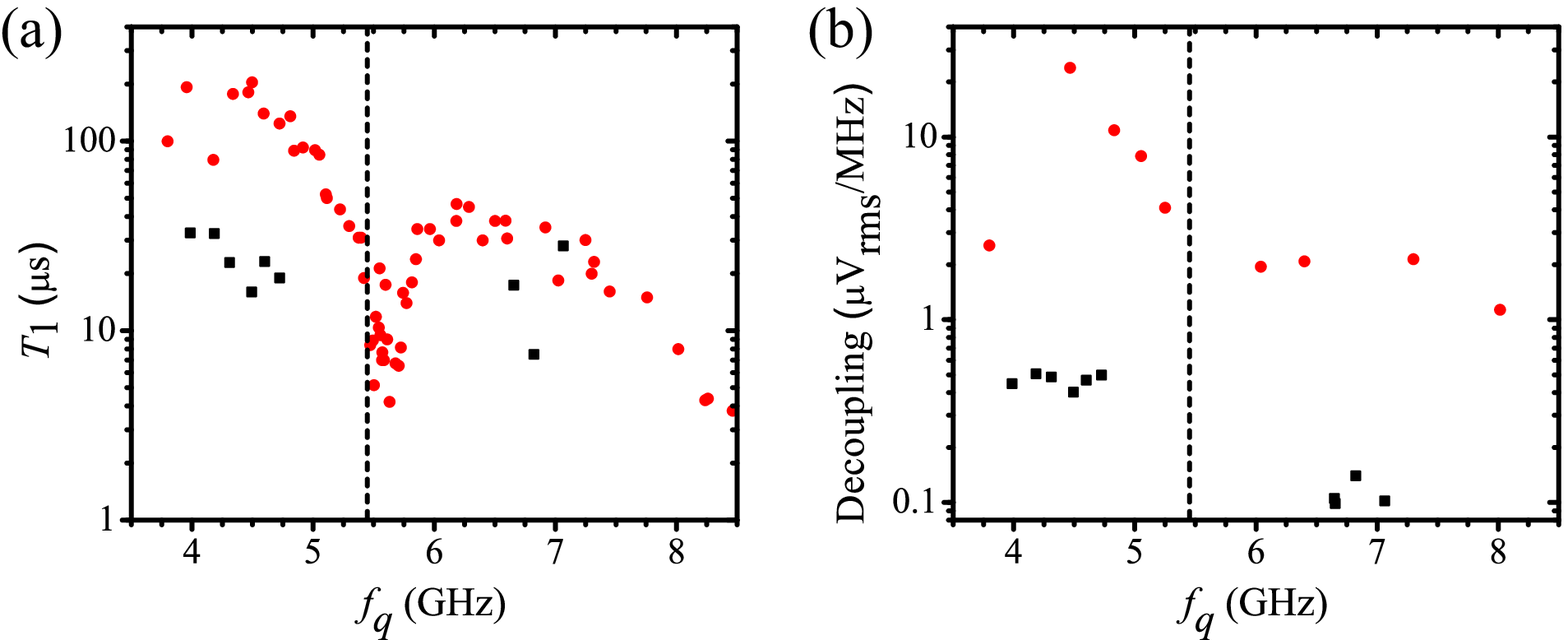}

\caption{(a) Energy relaxation lifetime $T_{1}$ versus transition frequency
$f_{q}$ and (b) qubit decoupling versus transition frequency $f_{q}$
for device 1 (\textcolor{red}{\CIRCLE{}}) and device 2 (\textifsymbol[ifgeo]{80}).
Lower decoupling correlates with lower $T_{1}$ for both devices.
The dashed vertical lines mark the resonator frequency $\unit[\approx5.45]{GHz}$.\label{fig:comparison}}
\end{figure*}
The qualitative behavior of $T_{1}$ in device 2 was similar to that
observed for device 1 {[}see Fig. \ref{fig:comparison}{]}. With the
qubit transition frequency $\omega_{q}$ biased below the resonator
$\omega_{r}$, the lifetime $T_{1}$ of device 2 was in the $\unit[15-30]{\lyxmathsym{\textgreek{m}}s}$
range while in device 1 $T_{1}$ was $\unit[30-200]{\lyxmathsym{\textgreek{m}}s}$.
In both devices there is a drop in $T_{1}$ for $\omega_{q}>\omega_{r}$.
For device 2, the decoupling $dV/df_{Rabi}$ was generally correlated
with $T_{1}$ and, in particular, both $T_{1}$ and $dV/df_{Rabi}$
drop for $\omega_{q}>\omega_{r}$. Above $\omega_{r}$, the lifetimes
for device 1 and device 2 are nearly identical while below $\omega_{r}$
the device 2 lifetimes are about a factor of 5 shorter than those
of device 1 {[}see Fig. \ref{fig:comparison}(a){]}. The decoupling
follows a similar qualitative trend {[}see Fig. \ref{fig:comparison}(b){]}.
We note that while the $T_{1}$ of device 2 is not as long as that
of device 1, it is still relatively long-lived for a superconducting
qubit and comparable to some 3D qubits.\cite{paik2011observation}

Some of the performance differences between the two samples can be
understood from small differences in the design and fabrication. Specifically
the larger Rabi $df_{Rabi}/dV$ coupling in device 2 may be due to
the fact that its resonator was more strongly coupled to the transmission
line. This is consistent with the smaller $Q_{e}$ in device 2. Additionally,
as seen from Eq. \ref{eq:chrgtrns}, the decoupling scales as $1/C_{g}E_{c}$.
Although device 2 had a somewhat smaller $E_{c}$ {[}see Table \ref{tab:device-summary}{]},
the CPB island was approximately four times longer and hence $C_{g}$
was correspondingly increased ($C_{g,1}=\unit[4.5]{aF}$ versus $C_{g,2}=\unit[19.1]{aF}$).
While the decoupling only addresses the decay channel due to the transmission
line, relaxation due to an interaction with discrete charged defects
may also be playing a role. Device 1 had no prominent splittings in
the transition spectrum while device 2 had visible splittings and
an anomalous multi-parabola spectrum {[}see Fig. \ref{fig:spectrum}(b){]}.
The larger area of the tunnel junctions of device 2 should have led
on average to about twice as many defects, which highlights the pitfalls
of trying to improve a CPB by using larger area junctions.

It is interesting to compare the spectral density of charge noise
at both low and high frequencies in the two devices. Assuming relaxation
due to charge noise dominates other decay channels, the spectral density
of charge noise $S_{Q}$ and the lifetime $T_{1}$ are related by
Eq. \ref{eq:sqandt1}. Using $T_{1}=\unit[16]{\text{\textgreek{m}}s}$
at $f_{q}=\unit[4.5]{GHz}$ for device 2 this places a bound on the
noise of $S_{Q}\left(f=\unit[4.5]{GHz}\right)\leq\unit[10^{-17}]{\textit{e}^{2}/Hz}$.
This is an order of magnitude larger than $S_{Q}\left(f=\unit[4.5]{GHz}\right)\leq\unit[10^{-18}]{\textit{e}^{2}/Hz}$
found at $f_{q}=\unit[4.5]{GHz}$ for device 1 but is similar to other
values reported in the literature.\cite{vion2002manipulating} The
spectroscopic coherence time $T_{2}^{*}$ of device 2, as determined
from low pump power spectral linewidth measurements, was at most $\unit[60]{ns}$.
This places a low frequency charge noise bound of $S_{Q}\left(f=\unit[1]{Hz}\right)\leq\unit[\left(1\times10^{-2}\right)^{2}]{\textit{e}^{2}/Hz}$
for device 2. This is an order of magnitude larger than the $T_{2}^{*}=\unit[200-500]{ns}$
found for device 1 from Ramsey measurements. Measurements of the decay
time of Rabi oscillations showed similar behavior: $T^{\prime}=\unit[1-2]{\text{\textgreek{m}}s}$
for device 1 and $T^{\prime}=\unit[0.2-1.8]{\lyxmathsym{\textgreek{m}}s}$
for device 2. These differences in charge noise may simply be due
to sample-to-sample statistical variations in the number of TLS's
active at low and high frequencies, and would be expected when few
TLS's are present on average.

\section*{Conclusion}

In summary we measured the spectrum, lifetime, Rabi oscillations,
Ramsey fringes, and spin echoes in two CPB charge qubits. The Ramsey
fringe decay times $T_{2}^{*}$ for device 1 were in the $\unit[200-500]{ns}$
range while the spin echo envelope decay times $T_{echo}$ were in
the $\unit[2.4-3.3]{\lyxmathsym{\textgreek{m}}s}$ range. The low
frequency $1/f$ noise power spectral density was bounded by $S_{Q}\left(f=\unit[1]{Hz}\right)\leq\unit[\left(3\times10^{-3}\right)^{2}]{\textit{e}^{2}/Hz}$,
a value which is consistent with other reports on charge qubits.\cite{wallraff2005approaching,siddiqi2006dispersive,vion2002manipulating,metcalfe2007measuring}
We also determined that the noise in device 1 is consistent with a
soft cutoff at $f_{c}\approx\unit[0.2]{MHz}$ and this is in turn
consistent with the exceptionally low bound on the high frequency
noise extracted from $T_{1}$ data. A second qubit based on the same
design also showed $T_{1}$ correlated to the decoupling $dV/df_{Rabi}$
in the transition frequency range $\unit[4.0-7.3]{GHz}$. The behavior
was qualitatively similar to that in device 1. The maximum lifetime
$T_{1}\approx\unit[30]{\text{\textgreek{m}}s}$ of device 2 was not
as long as the longest $T_{1}\approx\unit[200]{\text{\textgreek{m}}s}$
in device 1, but the coupling between the qubit and the transmission
line was $\approx10-20$ times stronger in device 2, suggesting some
level of reproducibility.
\begin{acknowledgments}
F. C. W. would like to acknowledge many useful discussions with K.
D. Osborn and C. J. Lobb and support from the Joint Quantum Institute
and the State of Maryland through the Center for Nanophysics and Advanced
Materials. V. Z. would like to thank M. Khalil, P. Nagornykh and N.
Siwak for many useful discussions.
\end{acknowledgments}
\bibliographystyle{aipnum4-1}

\end{document}